\RequirePackage{fix-cm}
\documentclass[authoryear,smallextended]{svjour3}       

\usepackage{graphicx}

\usepackage{amsmath}
\usepackage{amsfonts}
\usepackage{amssymb}
\usepackage{verbatim}
\usepackage{enumerate}
\usepackage[colorlinks,citecolor=blue,urlcolor=blue]{hyperref}
\usepackage[justification=centering]{caption}
\smartqed

\usepackage[12pt]{extsizes}

\usepackage{booktabs,caption}
\usepackage[flushleft]{threeparttable}

\spnewtheorem*{Main Theorem}{Main Theorem}{\normalfont\bfseries}{\itshape}
\spnewtheorem{mylemma}[theorem]{Lemma}{\bf}{\it}
\spnewtheorem{myproposition}[theorem]{Proposition}{\bf}{\it}
\spnewtheorem{mycorollary}[theorem]{Corollary}{\bf}{\it}
\spnewtheorem{mydefinition}[theorem]{Definition}{\bf}{\it}
\spnewtheorem{myquestion}{Question}{\bf}{\it}
\spnewtheorem{myconjecture}[myquestion]{Conjecture}{\bf}{\it}
\numberwithin{equation}{section} \numberwithin{theorem}{section}
\usepackage[utf8]{inputenc}
\usepackage{fourier} 
\usepackage{array}
\usepackage{makecell}
\usepackage{bm}

\usepackage{geometry}
\usepackage{natbib} 
\usepackage{filecontents}

\begin{document}

\title{Evolution of Regional Innovation with Spatial Knowledge Spillovers: Convergence or Divergence?}

\titlerunning{Evolution of Regional Innovation}        

\author{Jinwen Qiu \and Wenjian Liu \and Ning Ning
}


\institute{        
             Jinwen Qiu, Ning Ning \at
              Department of Statistics and Applied Probability, University
	of California, Santa Barbara\\
	\email{jqiu@pstat.ucsb.edu, ning@pstat.ucsb.edu} 
	\and
	Wenjian Liu \at
	Department of Mathematics and Computer Science,
	Queensborough Community College, City University of New York\\
	\email{wjliu@qcc.cuny.edu}   
}

\maketitle

\begin{abstract}
This paper extends endogenous economic growth models to incorporate knowledge externality. We explores whether spatial knowledge spillovers among regions exist, whether spatial knowledge spillovers promote regional innovative activities, and whether external knowledge spillovers affect the evolution of regional innovations in the long run. We empirically verify the theoretical results through applying spatial statistics and econometric model in the analysis of panel data of $31$ regions in China. An accurate estimate of the range of knowledge spillovers is achieved and the convergence of regional knowledge growth rate is found, with clear evidences that developing regions benefit more from external knowledge spillovers than developed regions.

\end{abstract}

\section{Introduction}
\label{intro}
Knowledge spillover is an exchange of ideas among individuals, which can be categorized as internal and external. Internal knowledge spillovers occur among individuals within an organization that produces goods and/or services, whereas external knowledge spillovers occur among individuals outside of a production organization. Knowledge spillovers are referred to non-rival knowledge market costs incurred by a party not agreeing to assume the costs that has a spillover effect of stimulating technological improvements in a neighbor through one's own innovation, in the language of knowledge management economics. According to the new trade theory and Endogenous Growth theory, technological progress is the only source to sustainable long-term economic growth (\cite{romer1986increasing}, \cite{grossman1993innovation}). From the history of global economic development, we can find that only the technological progress motivated by independent innovations can push up long-term economic growth in a sustainable way. In other words, economic development that relies on low labor cost and natural resources cannot last for a long period. Transforming extensive economic growth into technological intensive economic development is a significant step for the developing countries to become developed countries. More importantly, independent innovations play an indispensable role in this process. 

Technological progress not just relies on the input of research and development (R\&D) capital and research personnels, but also associates with the current knowledge level. \cite{jaffe1993geographic} stated that due to the localization feature of knowledge, it is difficult for knowledge to spread among different regions. However, based on a variety of research, for example \cite{krugman1991increasing}, \cite{griliches1991search}, \cite{coe1995international}, \cite{anselin2000geographical} and \cite{paci1999externalities}, we know that to some extent new knowledge is a kind of public goods and spillovers transcend political boundaries. Based on empirical studies in \cite{bottazzi2003innovation}, \cite{monjon2003assessing}, \cite{fischer2003spatial} and \cite{greunz2003geographically}, we know that knowledge spillovers had a remarkable impact on regional innovations, and the influence dwindles as the distance between one region and another increases. \cite{abdelmoula2007spatial} indicated that public R\&D has a stronger spillover effect than private R\&D, and knowledge spillovers from different regions carry two types of characteristics, which are ``competitiveness" and ``complementarity". \cite{ponds2009innovation} pointed out that the spatial scope cannot limit the establishment of official and unofficial cooperation network, and more importantly this network plays an indispensable role in the generation of knowledge spillovers among different regions. Various theoretical and empirical studies revealed the spatial dependence and agglomeration in regional innovative activities, as well as the positive impact of knowledge spillovers on regional innovations. It is well known that there are connections between regional knowledge spillovers and other factors such as geographical distance, learning capability and mechanism, regional culture, society, economy and politics. Therefore, knowledge spillovers exist to some extent in innovative activities regardless of industries, regions and countries; external knowledge spillovers also involuntarily boost the inner local innovative activities.

Furthermore, to increase national income and facilitate economic transition, narrowing the economic gap among different regions,  becomes an urgent concern to economists and governments as well as the societies, which is also a difficult problem to be solved. With the fast improvement of transportation infrastructure and the rapid development of tool for information transmission, knowledge communication and technology exchange become more and more frequent, making the phenomenon of knowledge spillovers more obvious. A recent and general example of a knowledge spillover is the collective growth associated with the research and development of online social networking tools such as Twitter, YouTube, Instagram and Facebook, which have not only created a positive feedback loop and a host of originally unintended benefits for their users, but have also created an explosion of new softwares, programming platforms, and conceptual breakthroughs that have perpetuated the development of the industry as a whole. 
Considering the close connections among different regions, to fulfill a strategic planning role and solve the urgent problem mentioned above, we need to answer the following questions: 1. Do spatial knowledge spillovers among regions exist? 2. Do spatial knowledge spillovers promote regional innovative activities? 3. What is the radiation range of spatial knowledge spillovers? 4. Do knowledge spillovers affect the evolution of regional innovations in the long run? To answer these questions, this paper thoroughly explored the underlying relations between spatial knowledge spillovers and regional innovations. 

Before displaying our result, it is worth mentioning some beautiful related literatures including, but not limited to, the following: For regional knowledge economies analysis, we refer to \cite{cooke2007regional, feser2003regions, todtling2006innovation, faggian2006human, bilbao2004r, paez2005spatial, capello2014spatial, andersson2006regional}. For knowledge spillovers analysis, we refer to \cite{anselin1997local, fischer2006geography, lesage2007knowledge, byun2005spillovers, breschi2010geography, bode2004spatial, duranton2000cumulative, rodriguez2008research}. For model learning including higher order spatial econometric modeling, we refer to \cite{partridge2010computable, irwin2010new, capello2005collective, elhorst2012model}. Analysis using Bayesian paradigm, we refer to \cite{lesage1997bayesian, lesage2007bayesian} and leave the machine learning in Bayesian setting (e.g. \cite{jammalamadaka2018multivariate}) to future possible extensions. For analysis specially for developing countries, we refer to \cite{grace2017spatial, fu2016reducing, fujita2004spatial}. For policy making which we concluded at the end of this paper, we refer to \cite{kinsella2010evaluating, lacombe2004does}.

The paper proceeds as follows.
Section \ref{Sec:Mechanism_Interpretation} establishes the dynamic casual relationship framework and proposes an innovative growth model for two regions. 
Section \ref{Sec:General_Innovative} generalizes the innovative growth model to $N$ regions and derives equilibrium solutions.
Section \ref{Sec:Spatial_Association} examines the spatial dependence of regional innovative activities based on spatial statistics. 
Section \ref{Sec:Analysis_of_Knowledge} empirically assesses the effect of knowledge spillovers on regional innovation. 
Section \ref{Sec:Knowledge_Spillover_Range} investigates the radiation range of knowledge spillovers in China.
Section \ref{Sec:Convergence_Property} explores convergence property of regional knowledge growth rate using empirical data.
Section \ref{Sec:Conclusion} concludes the paper and offers some policy suggestions.

\section{Mechanism Interpretation and Theoretical Model}
\label{Sec:Mechanism_Interpretation}
\subsection{Mechanism Interpretation}
 Successfully using knowledge spillovers to promote regional innovations, depends on knowledge owners' production motivation and capacity of new knowledge, as well as knowledge recipients' learning ability and production motivation of new knowledge. Firstly, from knowledge owners' perspective, on one hand, with improved production capacity of new knowledge, knowledge owners will be inclined to produce more new knowledge; on the other hand, decreased production cost of physical products will enhance its competitiveness in the market, which in return gives more profits to knowledge owners. Consequently, more funds will be invested in the production of new knowledge, forming a positive feedback loop. As knowledge owner's knowledge stock expands, three different types of knowledge spillovers will occur. The first type is described as involuntary knowledge spillovers, which means that knowledge recipients gain external knowledge through hiring researchers who use the same research methods as still used by knowledge owners. It seems that knowledge owners just do not obtain benefits in this process and even lose its incentives to produce new knowledge. However, using benefits from spillovers, knowledge recipients can reduce production cost to increase its competitiveness, which has a tremendous negative impact on knowledge owners' profit space. As a result, knowledge owners will be forced not only to strengthen its current protection of its knowledge stock but to invest more in knowledge production to win back its lost competitiveness. The second type is voluntary cooperative knowledge spillovers, whose definition is that knowledge owners and recipients do joint R\&D and can share a part of each other’s knowledge stock. Through cooperation, knowledge owners benefit not only from shared risks accompanied by R\&D but also from a part of complementary knowledge owned by recipients, both of which effectively stimulate knowledge owners' production motivation of new knowledge and then increase their knowledge stock. The third type is voluntary knowledge spillovers from transfer of knowledge, which refers that knowledge owners sell the newly created knowledge to recipients. In this process, owners are more willing to produce new knowledge because of the revenue from selling newly created knowledge.

\begin{figure}
	\includegraphics[scale=0.6]{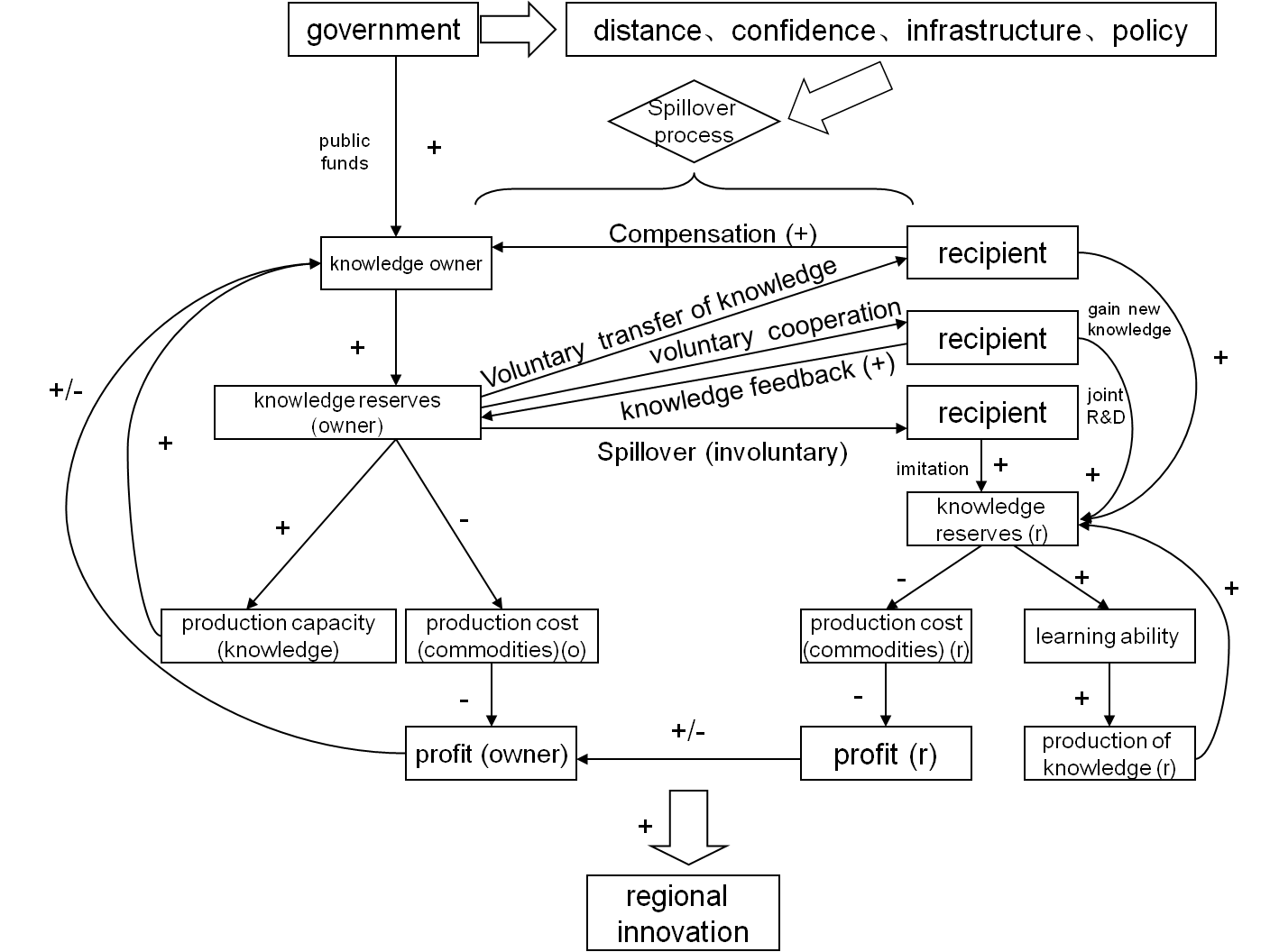}
	\caption{Dynamic Relationship between Knowledge Spillover and Regional Innovation}
	\label{fig:flowchart}       
\end{figure}

Secondly, from recipients' perspective, no matter what type of knowledge spillovers is, knowledge recipients will benefit from external knowledge spillovers, in increasing its knowledge stock and improving learning ability of new knowledge, both of which eventually promote its production capacity of new knowledge. Moreover, the dynamic process of knowledge spillovers is affected by a variety of factors such as spatial distance, confidence degree between knowledge owners and recipients, the condition of telecommunication and transportation infrastructure, relevant government policies, etc. Finally, based on the analysis above, knowledge spillovers greatly benefit regional innovations by enlarging knowledge stock of both knowledge owners and recipients. Therefore, to boost economic growth, the government should conditionally support knowledge spillovers through offering public funds to knowledge owners, establishing intermediary agencies to facilitate knowledge transfer, creating communicative and interactive platforms for different companies, encouraging industrial collaborations, all of which will further raise knowledge owners' enthusiasm of new knowledge production and eventually create a virtuous cycle of economic development driven by regional innovations. Figure \ref{fig:flowchart} indicates a dynamic feedback map on how knowledge spillovers propel regional innovations. Sign `+' indicates positive impact, sign `-' indicates negative impact and sign `+/-' indicates positive or negative impact.

\subsection{Innovative Growth Model with Knowledge Spillover for Two Regions}

\subsubsection{Model Assumption and Specification}

This paper extend endogenous economic growth models proposed by \cite{romer1990endogenous}, \cite{grossman1993innovation} through incorporating knowledge externality. Without loss of generality and for simplicity, we make the following three assumptions. Firstly, there are two regions in an open economy. Production functions of both industrial products and new knowledge are defined as general Cobb-Douglas production functions, where the sum of production inputs' parameters is not limited to one. Except parameters for exterior knowledge stock, all parameters of inputs for two regions are the same. Secondly, saving rate as well as the proportion of capital reserve and labor invested in knowledge production to that invested in goods production are set as exogenous and the same for two regions. Thirdly, capital and knowledge depreciation rate are assumed to be zero. Furthermore, interactive factors between two regions such as labor movement and trade are embodied by external knowledge spillover variables.

This model has two sectors and four variables. Two sectors are the industrial sector and the R\&D sector, with respective to output of industrial products and new knowledge; four variables are industrial products output ($Y$), capital reserve ($K$), labor force ($L$) and knowledge stock ($A$). We denote $(1-a_K)K$ and $a_K K$  as the capital input invested in the industrial sector and the R\&D sector respectively, and similar notations $(1-a_L)L$ and $a_L L$  for the labor input invested. Consequently the model describing two regions’ economic and innovative activities can be presented as:
\begin{equation}\label{eq:two_model_Y}
Y_i(t)=[(1-a_K)K_i(t)]^{\alpha}\cdot [A_i(t)(1-a_L)L_i(t)]^{1-\alpha}, \quad 0<\alpha<1,  i=1,2,
\end{equation}
\begin{equation}\label{eq:two_model_A1}
\dot{A}_1(t)=B[a_K K_1(t)]^{\beta} \cdot [a_L L_1(t)]^{\gamma}\cdot A_1(t)^{\theta}\cdot A_2(t)^{\mu_1}, \quad B>0, \beta \geq 0, \gamma\geq 0,
\end{equation}
\begin{equation}\label{eq:two_model_A_2}
\dot{A}_2(t)=B[a_K K_2(t)]^{\beta} \cdot [a_L L_2(t)]^{\gamma}\cdot A_2(t)^{\theta}\cdot A_1(t)^{\mu_2}, \quad B>0, \beta \geq 0, \gamma\geq 0,
\end{equation}
with $i$ representing region $i$. Parameter $\theta$ represents the elasticity of current knowledge stock in local region to new knowledge production. Parameter $\mu_1$ and $\mu_2$ respectively represents first region's and second region's elasticity of exterior knowledge stock to knowledge production, indicating the absorption and learning ability of exterior knowledge. The value of parameters should satisfy the condition $0\leq \mu_i<1-\theta-\beta$ for $i=1$ or $2$, so divergent knowledge growth in two regions is excluded. Moreover, in the process of knowledge production, knowledge stock in both local region and exterior region serve as input that reflects interactive connection between two regions' economic and innovative activities. Saving rate in two regions is set as exogenous and constant and depreciation rate is assumed to be zero:
\begin{equation}\label{eq:two_model_savingrate} 
\dot{K}_i(t)=s_iY_i(t), \quad s\geq 0, \quad i=1,2.
\end{equation}
Labor force growth rate in two regions are also set as exogenous and constant:
\begin{equation}\label{eq:two_model_growthrate} 
\dot{L}_i(t)=n_i L_i(t), \quad i=1,2.
\end{equation}

\subsubsection{Analysis of Knowledge and Capital Dynamics for Two Regions}
To begin with, equation \eqref{eq:two_model_A1} and \eqref{eq:two_model_A_2} are respectively divided by $A_1(t)$ and $A_2(t)$, on both side. Hence the growth rate of $A_1(t)$ and $A_2(t)$ can be expressed as:
\begin{equation}\label{eq:two_model_A1_growth} 
g_{A_1}(t)=c_AK_1(t)^{\beta}L_1(t)^{\gamma}A_1(t)^{\theta-1}A_2(t)^{\mu_1}
\end{equation}
\begin{equation}\label{eq:two_model_A_2_growth} 
g_{A_2}(t)=c_AK_2(t)^{\beta}L_2(t)^{\gamma}A_2(t)^{\theta-1}A_1(t)^{\mu_2}
\end{equation}
In the equation above, $c_A=Ba_K^{\beta}a_L^{\gamma}$. Taking logarithm of equations \eqref{eq:two_model_A1_growth} and \eqref{eq:two_model_A_2_growth} on both side and then differentiating them with respect to time $t$ yields:
\begin{equation}
\frac{\dot{g}_{A_1}(t)}{g_{A_1}(t)}=\beta g_{K1}(t)+\gamma n_1+(\theta-1)g_{A_1}(t)+\mu_1 g_{A_2}(t)
\end{equation}
\begin{equation}
\frac{\dot{g}_{A_2}(t)}{g_{A_2}(t)}=\beta g_{K2}(t)+\gamma n_2+(\theta-1)g_{A_2}(t)+\mu_2 g_{A_1}(t).
\end{equation}
For the first region, if $\dot{g}_{A_1}(t)>0$ , then $\beta g_{K1}(t)+\gamma n_1+(\theta-1)g_{A_1}(t)+\mu_1 g_{A_2}(t)$ is positive and $g_{A_1}(t)$ increases. Because $\theta-1$ is smaller than zero, $\beta g_{K1}(t)+\gamma n_1+(\theta-1)g_{A_1}(t)+\mu_1 g_{A_2}(t)$ decreases until the equation is equal to $0$; that is to say $\dot{g}_{A_1}(t)=0$ and $g_{A_1}(t)$ remains constant. If $\dot{g}_{A_1}(t)<0$, then $g_{A_1}(t)$ rises and hence $\beta g_{K1}(t)+\gamma n_1+(\theta-1)g_{A_1}(t)+\mu_1 g_{A_2}(t)$ gradually increases until the equation is equal to $0$; that is to say $\dot{g}_{A_1}(t)=0$ and remains constant. If $\dot{g}_{A_1}(t)$ is equal to zero, then $g_{A_1}(t)$ will keep constant. Similarly, the same conclusion can be drawn for the second region.

In addition, substitute equation \eqref{eq:two_model_Y} into equation \eqref{eq:two_model_savingrate}, and then is respectively divided by $K_i(t)$ for $i=1, 2$ on both sides. The growth rate of $K_i(t)$ can then be expressed as:
\begin{equation}\label{eq:two_model_growth_rate_K} 
g_{K_i}(t)=c_{Ki} \left ( \frac{A_i(t)L_i(t)}{K_i(t)} \right)^{1-\alpha}, \quad i=1, 2.
\end{equation}
In the model, $c_{Ki}=s_i(1-a_K)^{\alpha}(1-a_L)^{1-\alpha}$. Taking logarithm of equation \eqref{eq:two_model_growth_rate_K} on both side and then differentiating them with respect to time $t$, yields:
\begin{equation}
\frac{\dot{g}_{Ki}(t)}{g_{K_i}(t)}=(1-\alpha)[g_{A_i}(t)+n_i-g_{K_i}(t)], \quad i=1, 2.
\end{equation}
If $\dot{g}_{Ki}(t)>0$, then $g_{K_i}(t)$ increases and $g_{A_i}(t)+n_i-g_{K_i}(t)>0$. If $\dot{g}_{Ki}(t)<0$, then $g_{K_i}(t)$ decreases and $g_{A_i}(t)+n_i-g_{K_i}(t)<0$. If $\dot{g}_{Ki}(t)=0$, then $g_{K_i}(t)$ keeps constant. The analysis of capital dynamics is similar to that of knowledge dynamics. No matter what initial value of $g_{K_i}$ is, based on its dynamic equation it will converge to steady state; that is to say, $\dot{g}_{Ki}(t)=0$  and  $g_{K_i}(t)$ keeps constant. 

\subsubsection{Equilibrium Solutions and Comparative Analysis of Growth Rate}

According to endogenous economic growth models, if economic activities are independent and parameters of factor input are the same in different regions, knowledge and capital growth rate will tend toward and eventually settle at an equilibrium point. Does knowledge externality affect the conclusion made above? Based on the dynamic analysis above, the growth rates of two regions’ knowledge and goods production will converge to the their positive steady equilibrium points, no matter what their initial values are, i.e. $\dot{g}_{A_i}$ and $\dot{g}_{Ki}$ are equal to $0$ for $i=1, 2$. Therefore, four equilibrium solutions are given by:
\begin{equation}
g_{A_1}^{\ast}=\frac{(1-\theta-\beta)n_1+\mu_1n_2}{(1-\theta-\beta)^2-\mu_1 \mu_2}(\gamma+\beta),
\end{equation}
\begin{equation}
g_{A_2}^{\ast}=\frac{(1-\theta-\beta)n_2+\mu_2n_1}{(1-\theta-\beta)^2-\mu_1 \mu_2}(\gamma+\beta),
\end{equation}
\begin{equation}
g_{K1}^{\ast}=\frac{(1-\theta-\beta)n_1+\mu_1n_2}{(1-\theta-\beta)^2-\mu_1 \mu_2}(\gamma+\beta)+n_1,
\end{equation}
\begin{equation}
g_{K2}^{\ast}=\frac{(1-\theta-\beta)n_2+\mu_2n_1}{(1-\theta-\beta)^2-\mu_1 \mu_2}(\gamma+\beta)+n_2.
\end{equation}

We can see that, the steady growth rate of knowledge in local region not only depends on its population growth, but also is positively influenced by exterior region’s population growth. On one hand, as population grow rapidly in other regions, to search for job opportunities, more people are inclined to move from one region to other regions and hence will bring new knowledge into exterior regions. On the other hand, a larger population leads to more prevalent interactive communications among different regions, promoting new knowledge sharing and learning. As a result, researchers in a local region will be inspired by knowledge from exterior regions and then produce more new knowledge; in other words, it will promote growth rate of knowledge. It conforms to reality in the activities of regional innovations . 

Furthermore, we analyze the absorption and learning ability of exterior knowledge, i.e. the effect of $\mu_1$ and $\mu_2$ on the steady growth rate of knowledge. Under the conditions that $n_1=n_2$ and $\mu_1=\mu_2$, the growth rate of knowledge for two regions are the same, that is $g_{A_1}^{\ast}=g_{A_2}^{\ast}$, which is consistent with results derived from existing models without knowledge spillover. It indicates that given other exogenous variables and the same parameters of other factor inputs, the growth rate of knowledge will eventually tends toward an equilibrium state in the long run, which contradicts the occurrence of accumulative cyclical effects. In other words, the region with low knowledge stock will catch up with the region with high knowledge stock in terms of growth rate of knowledge or innovative ability. Under the conditions that $n_1=n_2$ and $\mu_1>\mu_2$, the first region’s growth rate of knowledge in equilibrium is greater than that of the second region, that is $g_{A_1}^{\ast}>g_{A_2}^{\ast}$, which shows that the absorption and learning ability of exterior knowledge plays a significant role in enhancing steady growth rate of knowledge in the local region.

\section{General Innovative Growth Model with Knowledge Spillover}
\label{Sec:General_Innovative}
\subsection{General Innovative Growth Model Setup}
Through the innovative growth model with two regions, given the same parameters and exogenous variables, we have shown that the growth rate of knowledge in two regions will eventually tends toward the same steady state in the long run. A natural question: If there are $N$ regions involved instead of just two, will the same conclusion hold? To answer this question, firstly we analyze how different regions are economically correlated by knowledge spillover. Well known, population movements, interactive communications and cooperations, and capital trades will lead to knowledge spillovers. In this paper, we consider the case that different regions exist in one economic entity and hence their economic connections are much closer, since the influence from trade barrier or cultural difference is not obvious. The most important factor that affects the degree of knowledge spillovers, is the spatial distance between two regions. For example, when one region is closer to another in terms of spatial distance, population flow and capital trade between these two regions are likely to be more prevalent. Consequently, knowledge stock in one region has a greater impact on the other’s innovative activities, that is to say, there are higher intensity of knowledge spillovers between these two regions. This idea is reflected in the general innovative growth model with knowledge spillovers.

Secondly, in this setting with $N$ correlated regions, without loss of generality, we make assumptions for simplicity: the population growth rate $n$ is the same for all regions, and the elasticity of exterior knowledge stock to local knowledge production $\mu$ is the same for all regions. Therefore, the model describing $N$ regions’ economic and innovative activities can be presented as:
\begin{equation}\label{eq:Nregions_Y}
Y_i(t)=[(1-a_K)K_i(t)]^{\alpha}\cdot [A_i(t)(1-a_L)L_i(t)]^{1-\alpha}, \quad 0<\alpha<1, i=1,2\cdots,N,
\end{equation}
\begin{equation}
\dot{A}_i(t)=B[a_K K_i(t)]^{\beta} \cdot [a_L L_i(t)]^{\gamma}\cdot A_i(t)^{\theta}\cdot \prod_{j=1}^N A_j(t)^{\mu w_{ij}}, \quad B>0, \beta \geq 0, \gamma\geq 0, 0\leq \theta,  \mu<1.
\end{equation}
As before, we set $\beta+\theta+\mu<1$ to guarantee the convergence of the growth of knowledge. Parameter $\mu$ describes the absorption and learning ability of knowledge in all other exterior regions. $w_{ij}$ contributes to a geometric weighted average of knowledge stock in all other exterior regions and reflects the relative connection between region $i$ and region $j$. For example, a greater $w_{ij}$ indicates a shorter spatial distance between two regions $i$ and $j$, a stronger relative connection as well. Moreover, we assume $w_{ij}$ is a non-negative constant variable, such that $0\leq w_{ij}\leq 1$, $w_{ii}=0$ and $\sum_{j=1}^Nw_{ij}=1$.  
Saving rates in $N$ regions are set as exogenous and the same:
\begin{equation}\label{eq:N_savingrate}
\dot{K}_i(t)=sY_i(t).
\end{equation}
Labor force growth rates in $N$ regions are also set as exogenous and the same:
\begin{equation}
\dot{L}_i(t)=nL_i(t).
\end{equation}

\subsection{Analysis of Knowledge and Capital Dynamics for $N$ Regions}
The growth rate of $A_i(t)$ can be achieved by dividing $A_i(t)$ on equation \eqref{eq:Nregions_Y}: 
\begin{equation}
g_{A_i}(t)=c_A [K_i(t)]^{\beta} \cdot [L_i(t)]^{\gamma}\cdot A_i(t)^{\theta-1}\cdot \prod_{j=1}^N A_j(t)^{\mu w_{ij}}, \quad c_A=Ba_K^{\beta}a_{L}^{\gamma}.
\end{equation}
Taking logarithm and then differentiating the above equation with respect to time $t$ yields:
\begin{equation}\label{eq:N_growthrate}
\frac{\dot{g}_{A_i}(t)}{g_{A_i}(t)}={\beta} g_{K_i}(t)+\gamma n+(\theta-1)g_{A_i}(t)+\mu \sum_{j=1}^N w_{ij}g_{A_j}(t), \quad i=1,\cdots,N.
\end{equation}
We see that, if $\dot{g}_{A_i}(t)>0$, growth rate $g_{A_i}(t)$ increases and 
$${\beta} g_{K_i}(t)+\gamma n+(\theta-1)g_{A_i}(t)+\mu \sum_{j=1}^N w_{ij}g_{A_j}(t)>0;$$ if $\dot{g}_{A_i}(t)<0$, then $g_{A_i}(t)$ decreases; if $\dot{g}_{A_i}(t)=0$, then $g_{A_i}(t)$ keep constant. Note that $\theta-1<0$, therefore when $g_{A_i}(t)$ increases, given other factors unchanged, we see that $${\beta} g_{K_i}(t)+\gamma n+(\theta-1)g_{A_i}(t)+\mu \sum_{j=1}^N w_{ij}g_{A_j}(t)$$ decreases and eventually goes to zero, which makes $\dot{g}_{A_i}(t)$ go to zero and $g_{A_i}(t)$ stay constant. Also, when we consider \eqref{eq:N_growthrate} with index $j$, that is
\begin{equation}
\frac{\dot{g}_{A_j}(t)}{g_{A_j}(t)}={\beta} g_{K_j}(t)+\gamma n+(\theta-1)g_{A_j}(t)+\mu \sum_{k=1,k\neq i}^N w_{jk}g_{A_k}(t)+\mu w_{ji}g_{A_i}(t),
\end{equation}
we see that, when $g_{A_i}(t)$ increases, given other factors unchanged, we see that $\frac{\dot{g}_{A_j}(t)}{g_{A_j}(t)}$ increase as $g_{A_i}(t)$ increase.
The influence of one region’s growth rate on another region's depends on the degree of connection ($w_{ij}$) between these two regions. 
Similarly, if $g_{A_i}(t)$  decreases, it will force the negative $\dot{g}_{A_i}(t)$ to go up while shrink down $\dot{g}_{A_j}(t)$ for $j=1,\cdots,i-1, i+1, \cdots, N$. As a result, the equation \eqref{eq:N_growthrate} implies that despite the existence of knowledge spillovers among $N$ different regions, growth rates of knowledge in different regions will tend toward steady states in the long run, which is consistent with the corresponding results in the cumulative cyclic causation theory.

In addition, after plugging equation \eqref{eq:Nregions_Y} into equation \eqref{eq:N_savingrate}, and dividing by $K_i(t)$ for $i=1,\cdots,N$ on both sides, the growth rate of $K_i(t)$ can be expressed as:
\begin{equation}
g_{K_i}(t)=c_K \left ( \frac{A_i(t)L_i(t)}{K_i(t)} \right)^{1-\alpha}, \quad c_k=s(1-a_K)^{\alpha}(1-a_L)^{1-\alpha}.
\end{equation}
Taking logarithm and differentiating them with respect to time $t$, yields:
\begin{equation}\label{eq:N_laborrate}
\frac{\dot{g}_{Ki}(t)}{g_{K_i}(t)}=(1-\alpha)[g_{A_i}(t)+n-g_{K_i}(t)], \quad i=1, \cdots, N.
\end{equation}
From equation \eqref{eq:N_growthrate} and \eqref{eq:N_laborrate}, we see that connections $w_{jk}$ among different regions is reflected through external knowledge terms and not reflected in external capital terms, that is, change of capital growth rate in local region is only associated with its own capital and knowledge growth rate. Similar analysis as the case with two regions before yields, no matter what the initial value of $g_{K_i}(t)$ is, it will eventually become stable, i.e. $\dot{g}_{Ki}(t)=0$ and $g_{K_i}(t)$ becomes a constant.

\subsection{Equilibrium Solutions and Analysis of Growth Rates in Equilibrium}
Based on the analysis above, $N$ different regions’ growth rates of knowledge and capital will transfer from unstable states to stable states. To check whether their growth rates are the same or related to $w_{ij}$,  we start with the equilibrium conditions that  $\dot{g}_{Ki}(t)=0$ and $\dot{g}_{A_i}(t)=0$ for $i=1,2,\cdots, N$, which provides $2N$ linear equations:
\begin{equation}
g_{K_i}^{\ast}(t)-g_{A_i}^{\ast}(t)=n, \quad i=1, \cdots, N,
\end{equation}
\begin{equation}
-\beta g_{K_i}^{\ast}(t)+(1-\theta)g_{A_i}^{\ast}(t)-\mu\sum_{j=1}^N w_{ij}g_{A_j}^{\ast}(t)=\gamma n.
\end{equation}
Note that $N$ different regions' interaction are modeled through $w_{ij}$, to reflect their related innovative and economic activities, we analyze the above equations as an interactive system instead of isolated linear equations. Therefore, we rewrite them in matrix form, simply,
\begin{equation}
\left( \begin{array}{cc} I & -I \\
-\beta I & (1-\theta)I-\mu W \end{array} \right)  
\left( \begin{array}{c} G_K \\
G_A \end{array} \right) =
\left( \begin{array}{c} \tilde{n} \\
\gamma \tilde{n} \end{array} \right).
\end{equation}
Here, $I$ represent the $N\times N$ identity matrix; $W$ represents the matrix $(w_{ij})_{N\times N}$ ; $G_K$, 
$G_A$ and $\tilde{n}$ are $(N\times 1)$ vectors, representing $(g_{K_1}^{\ast}, g_{K_2}^{\ast}, \cdots, g_{K_N}^{\ast})^{T}$, $(g_{A_1}^{\ast}, g_{A_2}^{\ast}, \cdots, g_{A_N}^{\ast})^{T}$ and $(n,n,\cdots,n)^{T}$ respectively. To solve this linear Simultaneous equation, use elementary operation on augmented matrix:
\begin{equation}
\left( \begin{array}{cc|c} I \quad \quad & -I  \quad\quad  & \quad \tilde{n} \\
-\beta I \quad \quad & (1-\theta)I-\mu W\quad \quad &\quad  \gamma \tilde{n} \end{array} \right) \rightarrow 
\left( \begin{array}{cc|c} I\quad \quad & -I \quad \quad & \quad  \tilde{n} \\
0 \quad \quad& (1-\theta-\beta)I-\mu W \quad \quad& \quad (\gamma+\beta) \tilde{n} \end{array} \right).
\end{equation}
Therefore, we have
\begin{equation}
\left( I-\frac{\mu}{1-\theta-\beta}W\right)G_A=\frac{\gamma+\beta}{1-\theta-\beta}\tilde{n}.
\end{equation}
Given $\beta+\theta+\mu<1$ and the properties of the $W$ matrix, we know that $\left | I-\frac{\mu}{1-\theta-\beta}W\right |\neq 0$ and the inverse matrix of $I-\frac{\mu}{1-\theta-\beta}W$ exists. Therefore, through analyzing each element in the matrix
\begin{equation}
G_A=\frac{\gamma+\beta}{1-\theta-\beta}\left( I-\frac{\mu}{1-\theta-\beta}W\right)^{-1}\tilde{n},
\end{equation}
we have
\begin{equation}
g_{A_i}^{\ast}=\frac{(\gamma+\beta)n}{1-\theta-\beta}\left( 1+\sum_{r=1}^{\infty}\left(\frac{\mu}{1-\theta-\beta}\right)^r \sum_{j=1}^N w_{ij}^{(r)}\right), \quad i=1,\cdots,N.
\end{equation}
The element $w_{ij}^{(r)}$ refers to the $r$th power of the entry in the $i$th row and $j$th column of the matrix $W$. Note that, no matter how many power the $W$ matrix takes, it still satisfies the property that $\sum_{j=1}^N w_{ij}^{(1)}=\sum_{j=1}^N w_{ij}^{(2)}=\cdots=1$, for $i=1,\cdots,N$. Moreover, given the condition that $1-\theta-\beta-\mu>0$, the solutions can be rewritten as:
\begin{equation}
g_{A_i}^{\ast}=\frac{(\gamma+\beta)n}{1-\theta-\beta-\mu}, \quad 
g_{K_i}^{\ast}=\frac{(1-\theta-\mu+\gamma)n}{1-\theta-\beta-\mu}, \quad i=1,\cdots,N.
\end{equation}
From the equilibrium solutions above, we see that each parameter is positively correlated with the steady growth rate of knowledge and capital. When a parameter changes, for example, if the absorption and learning ability of exterior knowledge parameter $\mu$ is improved, the steady growth rate will rise. Another significant conclusion is that for an economic entity with $N$  economically connected regions with knowledge spillovers, given the same parameters as well as exogenous variables, $N$ different regions’ growth rates of knowledge and capital will tend toward stable states and are independent of the level of association among $N$ regions, i.e. $w_{ij}$. If the factor endowment difference among $N$ regions is not obvious, the gap of the innovative capacity among $N$ regions would gradually narrow down in the long run, leading to the convergence of regional knowledge output, which is consistent with the results from neoclassic regional knowledge diffusion theory. Also, the results derived from the theoretical models above reflect an evolutionary principle of regional innovative activities to some extent. 

\section{Spatial Association of Regional Innovative Activities}
\label{Sec:Spatial_Association}
\begin{figure}
	\includegraphics[]{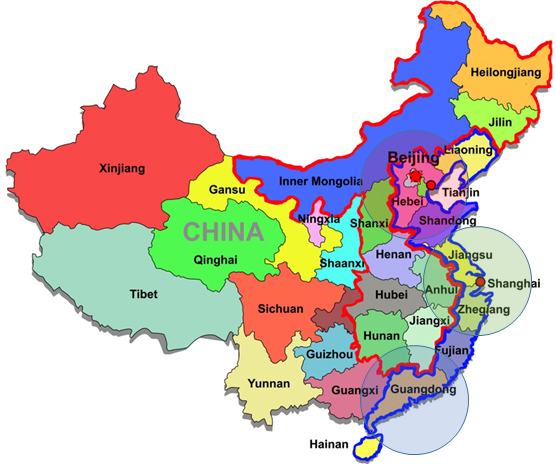}
	\caption{Map of China. The regions covered by bold red (resp. blue) line is the central (resp. east) area and the rest is categorized as west area. The circles are the radiation ranges of $1000$ kilometers centered at Beijing, Shanghai and Guangzhou, respectively.}
	\label{fig:china}       
\end{figure}

In this section, we empirically verify the theoretical results through applying spatial statistics and econometric model in the analysis of panel data of $31$ regions in China (a map of China showing provinces (regions) can be seen in Figure \ref{fig:china}. )
That is, we employed exploratory spatial data analysis to study spatial association of regional innovative activities and then revealed the existence of spatial knowledge spillovers. 

\subsection{Spatial Weighted Matrix}
In order to effectively identify the interaction among a set of spatial units, we quantify the level of the spatial interaction. In the sequel, we constructed a spatial weighted matrix, where the  entries show the strength of spatial interaction between regions which are set exogenously. 

Customarily, the strength of spatial interaction among spatial units is assumed to be geographic proximity, which means that only adjacent regions’ observable variables have spatial interaction  and hence a binary spatial weighted matrix is generated. However, there are two main drawbacks for this kind of matrix. Firstly, the assumption about geographic proximity is not realistic. Even if one region is not adjacent to or is far away from another region, spatial interaction between two regions still exists to some extent though it might be very week. Secondly, the binary spatial weighted matrix cannot accurately describe the relative effect of other regions on one specific region. Therefore, inspired by the distance-decay function of Geographical Gravity Model, this paper defines elements in spatial weighted matrix to be the reciprocal of square of distance between two regions, that is
\begin{equation}
w_{ij}^{\ast} =     \left\{ \begin{array}{rcl}
        &0 & \mbox{for }
          i=j\\ 
         &\frac{1}{d_{ij}^2} & \mbox{for }  i \neq j
                \end{array}\right..
\end{equation}
Here, the entry $d_{ij}$ represents the spatial distance between region $i$ and
region $j$, which is defined as the spatial geographic distance or spatial time distance. This paper measures the spatial geographic distance $w_g$ and spatial time distance $w_{t^2}$ between provincial capital cities in China via navigation map to represent spatial distance between two regions. Furthermore, this paper standardized the $W$ matrix, that is to say $w_{ij}=\frac{w_{ij}^{\ast}}{\sum_{i=1}^Nw_{ij}^{\ast}}$, where $N$ is the number of regions. 

\subsection{Global Spatial Association of Innovative Activities}
We study the global spatial association of regional innovative activities by a Spatial Autocorrelation tool, Global Moran's $I$ statistic, which measures spatial autocorrelation based on both feature locations and feature values simultaneously and characterizes the clustering of regional spatial distribution as a whole. 
That is, given a set of features and an associated attribute, it evaluates whether the pattern expressed is clustered, dispersed, or random. The Moran's $I$ is defined as follow:
\begin{equation}
I=\frac{\sum_{i=1}^n\sum_{j=1}^nw_{ij}(Y_{i}-\bar{Y})(Y_{j}-\bar{Y})}{S^2\sum_{i=1}^n\sum_{j=1}^nw_{ij}},
\end{equation}
$$S^2=\frac{1}{n}\sum_{i=1}^n(Y_i-\bar{Y})^2, \quad \bar{Y}=\frac{1}{n}\sum_{i=1}^n Y_i,$$
where $Y_i$ refers to the amount of authorized patents per ten thousand people in the paper in region $i$, which is observable, $n$ represents the number of total regions, and $w_{ij}$ is the element in spatial weighted matrix $W$. Here we need to emphasize that we use spatial geographic distance weighted matrix $W_g$ introduced above to calculate $w_{ij}$ in the Moran’s $I$ statistic. Moreover, the absolute value of the Moran’s $I$ statistic is set to be smaller than one. When Moran’s $I$ statistic is close to zero, it indicates the weak spatial association or no spatial association. When the Moran’s $I$ statistic is near one or negative one, it indicates strong positive or negative correlation among regions. Table \ref{table:Moran_I} indicates the Moran’s $I$ statistic of innovative activities for $31$ regions in China between $2000$ and $2016$. Since $2007$, all $p$ values for the Moran’s $I$ statistic is less than $10\%$, showing positive global spatial association of regional innovative activities. The spatial distribution of regional innovative activities shows obvious characteristics of spatial dependence and is not a random distribution, both of which further reveal the existence of spatial knowledge spillovers. In general, the value of Moran’s $I$ statistic gradually rises year by year, indicating stronger spatial association and spatial knowledge spillovers become more obvious. 
\begin{table}[h]
	\centering
	\centering
	\setlength{\tabcolsep}{8pt}
	\small
	\begin{tabular}{ccc}
		\hline\hline
		Year & Morgan's $I$ & $p$-value \\
		\hline
		 2000&0.009&0.584\\
         2001&0.019&0.478\\
         2002&0.029&0.378\\
         2003&0.109&0.067\\
         2004&0.043&0.297\\
         2005&0.056&0.226\\
         2006&0.096&0.107\\
         2007&0.121&0.062\\
         2008&0.118&0.073\\
         2009&0.146&0.034\\
         2010&0.147&0.031\\
         2011&0.165&0.013\\
         2012&0.159&0.014\\
         2013&0.155&0.022\\
         2014&0.148&0.031\\
         2015&0.132&0.050\\
		\hline
	\end{tabular}
	
    \caption{Moran’s I Statistic of Innovative Activities \label{table:Moran_I}}
\end{table}

\section{Analysis of Knowledge Spillover Effect}
\label{Sec:Analysis_of_Knowledge}
\subsection{Econometric Strategy and Data Sources}
\subsubsection{Econometric Model}
According to the general innovative growth model with knowledge spillovers, the knowledge production function can be expressed as follows:
\begin{equation}
\label{eq:dot_A}
\dot{A}_i(t)=B[a_K K_i(t)]^{\beta} \cdot [a_L L_i(t)]^{\gamma}\cdot A_i(t)^{\theta}\cdot \prod_{j=1}^N A_j(t)^{\mu w_{ij}}, \quad B>0, \beta \geq 0, \gamma\geq 0, 0\leq \theta,  \mu<1.
\end{equation}
In the process of both material production and knowledge production, inefficient production exists because of exterior shocks and interior factors. Simply put, given specific factors inputs, it is not sure for producers to achieve optimal outputs. Therefore, stochastic production frontier model was introduced by \cite{aigner1977formulation}. Through incorporating inefficiency terms into traditional production functions, the model solved problem of inaccurate estimates from traditional production functions.
Suppose in a world without error or inefficiency, at time $t$, the $i$th producer would produce:
$$Y_{i}(t)=F(X_{i}(t),\vartheta),$$  
where $X_i(t)$ represents the inputs such as $K_i(t)$ and $L_i(t)$ in equation \eqref{eq:dot_A},  $\vartheta$ represents the corresponding parameters such as $\beta$ and $\gamma$ in equation \eqref{eq:dot_A}, and function $F$ takes the form 
$$F((x_{i1},\cdots,x_{ik}),(y_{i1},\cdots, y_{ik}))=\sum_{j=1}^kx_{ij}^{y_{ij}}.$$

A fundamental element of stochastic frontier analysis is that each firm potentially produces less than it might because of a degree of inefficiency. Specifically,
$$Y_{i}(t)=F(X_{i}(t),\vartheta)\cdot \epsilon_{i}(t),$$   
where $\epsilon_{i}(t)\in (0,1]$ is the level of efficiency for producer $i$ at time $t$. If $\epsilon_{i}(t)=1$, the firm is achieving the optimal output with the technology embodied in the production function. When $\epsilon_{i}(t)<1$, the firm is not making the most of the inputs  given the technology embodied in the production function. Because the output is assumed to be strictly positive, the degree of technical efficiency is assumed to be strictly positive, that is $\epsilon_{i}(t)>0$. The output is assumed subject to random shocks, that is
$$Y_{i}(t)=F(X_{i}(t),\vartheta)\cdot \epsilon_{i}(t)\cdot \exp(v_{i}(t)),$$ 
with $v_{i}(t)\sim N(0,\sigma_v^2)$.  
Taking the natural logarithm of both sides and assuming that there are $k$ inputs, yields:
$$\ln Y_{i}(t)=\beta_0+\sum_{j=1}^k \beta_j \ln(x_{ij}(t))+v_{i}(t)-\mu_{i}(t),$$  
where  $x_{ij}(t)$ is a collection of production inputs
and $\mu_{i}(t)=-\ln \epsilon_{i}(t)$ such that $\mu_{i}(t)\geq 0$. There are two types of stochastic frontier models, the time-invariant model and the time-varying decay model. We adopt the time-varying decay stochastic frontier model, which is more prevalent in reality, and the inefficiency effects are modeled through
$$\mu_{i}(t)=\exp(-\eta (t-T_i))\mu_i.$$
Here, $T_i$ represents the last period in the $i$th panel; $\eta$ is the decay parameter; $\mu_i\sim N^+(\mu,\sigma_{\mu}^2)$ denotes the truncated-normal distribution, which is truncated at zero with mean $\mu$ and variance $\sigma_{\mu}^2$; $\mu_i$ and $v_{i}(t)$ are distributed independently of each other and the covariates in the model. When $\eta>0$ (resp. $\eta<0$), the degree of inefficiency decreases (resp. increases) over time and the level of inefficiency decays (resp. increases) toward the base level.

By combining the general innovative growth model with stochastic production frontier model, we construct a panel data econometric model as follow:
$$
\ln \dot{A}_i(t)=\beta_0+\beta_1 \ln K_{r,i}(t)+\beta_2 \ln L_{r,i}(t)+\beta_3 \ln A_i(t)+\beta_4 W \ln A(t)+v_{i}(t)-\mu_{i}(t).
$$
Here, $K_{r,i}(t)$ denotes the capital allocated to R\&D for region $i$ at time $t$, where $r$ stands for the proportion for research; $K_{r,i}(t)$ is $a_K K_i(t)$ in the general innovative growth model. $L_{r,i}(t)$ is $a_L L_i(t)$ in the general innovative growth model, representing R\&D personnel. $\ln A(t)$ represents $(\ln A_1(t), \ln A_2(t), \cdots, \ln A_n(t))^T$, where $n$ is the total number of regions. $W$ represents the spatial weighted matrix created above in the form of $W_t$ or $W_g$. The interactive term $W\ln A(t)$ represents knowledge spillovers.

\subsubsection{Data Sources}
All the data in this paper is collected from \textit{Chinese Science and Technology Statistical Yearbook and Chinese Statistical Yearbook} from $2000$ to $2016$.
\begin{itemize}
	\item \textbf{The number of authorized patents $P$ and the number of R\&D personnel $L$}\\
 The number of authorized patents, the number of scientific research papers which can be retrieved, and the sales revenue of new products can be categorized as knowledge productions. Considering the integrity and availability of data, we choose the number of authorized patents $P$ as the representative index for knowledge outputs. New knowledge outputs in period $t$ becomes authorized patents in period $t+1$, and $\dot{A}_i(t)$ is replaced by $P_i(t+1)$ in this model. We measure the unit of authorized patent as one and the unit of R\&D personnel as one person.

    \item \textbf{The R\&D capital reserve $K$ and the knowledge stock $A$}\\
The R\&D reserve in different regions is estimated by the perpetual inventory method and the formula is as follows:
	$$K_{r,i}(t)=(1-\delta)K_{r,i}(t-1)+R(t),$$ 
where the rate of depreciation of R\&D capital reserve $\delta$ is usually $10\%$ in China, which is the actual R\&D expense for each region per year and is converted into the R\&D inputs measured by purchasing power in $2000$ through the GDP deflater index. The R\&D capital reserve in different regions in $2000$ is calculated with the method proposed by \cite{griliches1991search}, that is $K_{r,2000}=\frac{R_{2000}}{g_i+\delta}$, where  is the geometric average growth rate of R\&D expense in different regions per year from $2000$ to $2011$, the unit of R\&D expense is $10,000$ Yuan. The knowledge stock can be calculated in the same way. Firstly we estimate knowledge stock in $2000$ namely $A_{2000}=\frac{P_{2000}}{g_i+\tau}$, where  is the geometric average growth rate of $P$ in each region from $2000$ to $2016$ and the depreciation rate of knowledge $\tau$ is $0.0714$, that is the reciprocal of the average service life of new knowledge in China ($14$ years). And then we are able to obtain knowledge stock in different regions per year by the perpetual inventory method.
\end{itemize}

\subsection{Impact of Knowledge Spillovers on Regional Innovations}

The disturbance term in the stochastic frontier model is assumed to have two components. One component is assumed to have a strictly nonnegative distribution, and the other is assumed to have a symmetric distribution. Therefore, only by using maximum likelihood method instead of ordinary least square method, we can obtain efficient and unbiased estimates. Furthermore, since China is a vast country and then regional endowment in disparate areas differs greatly, the same factors inputs will have different effects on regional knowledge productions. In view of geographical locations and development levels, this paper divide all regions into three groups, namely eastern, central and western areas for further comparative analysis. Fitting the regression model constructed above by means of two different spatial weighted matrices, yields estimation outputs in Table \ref{table:MLE_time_distance} and Table \ref{table:MLE_geographic_distance} respectively.

\begin{table}[h]
	\centering
	\centering
	\setlength{\tabcolsep}{8pt}
	\small
	\begin{tabular}{ccccc}
		\hline\hline
		\thead{Explanatory\\ variables} & \thead{Whole nation\\ ML estimate} & \thead{Eastern area\\ ML estimate} & \thead{Central area\\ ML estimate} & \thead{Western area\\ ML estimate}\\
		\hline
		lnKr &
0.0921*
(0.0554) &
0.2504**
(0.1096) &
-0.0517
(0.0769)&
-0.0392
(0.1095)\\
lnLr&
0.2418***
(0.0540)&
0.1691***
(0.0611)&
0.4103***
(0.0824)&
0.4366***
(0.1453)\\
lnA&
0.7477***
(0.0582)&
0.8077***
(0.1028)&
1.1015***
(0.0567)&
0.5957***
(0.1465)\\
wlnA&
0.1377***
(0.0466)&
-0.0541
(0.0839)&
0.2726***
(0.0381)&
0.1589**
(0.0687)\\
Constant term&
0.0153
(5.0317)&
-2.2387
(3.1415)&
-8.1695***
(0.6752)&
-0.2755
(2.9341)\\
$\mu_E$&
3.6702
(4.9204)&
1.7121
(2.3321)&
-0.1878
(1.0999)&
2.0617
(2.4800)\\
$\eta$&
0.0107
(4.9204)&
0.0181
(0.0184)&
0.0194
(0.0239)&
0.0292
(0.0272)\\
		\hline
	\end{tabular}
	\caption{Maximum Likelihood Estimates for Knowledge Spillover (time distance)\label{table:MLE_time_distance}}
		\begin{tablenotes}
		\centering
		\small
		\item Signif. codes:  0 ‘***’ 0.001 ‘**’ 0.01 ‘*’ 0.05 ‘.’ 0.1 ‘ ’ 1
		\item (numbers in brackets are standard deviation of the regression coefficient)
	\end{tablenotes}
\end{table}

\begin{table}[h]
	\centering
	\centering
	\setlength{\tabcolsep}{8pt}
	\small
	\begin{tabular}{ccccc}
		\hline\hline
		\thead{Explanatory\\ variables} & \thead{Whole nation\\ ML estimate} & \thead{Eastern area\\ ML estimate} & \thead{Central area\\ ML estimate} & \thead{Western area\\ ML estimate}\\
		\hline	
		lnKr&
0.1041*
(0.0559)&
0.2868**
(0.1250)&
-0.0671
(0.0858)&
0.0021
(0.1186)\\
lnLr &
0.2290***
(0.0533)&
0.1611***
(0.0629)&
0.4387***
(0.1092)&
0.4175***
(0.1394)\\
lnA &
0.7541***
(0.0582)&
0.7691***
(0.1079)&
1.0302***
(0.0591)&
0.5812***
(0.1199)\\
wlnA &
0.2026***
(0.0781)&
0.0587
(0.0841)&
0.3482***
(0.0634)&
0.2465*
(0.1532)\\
Constant term &
-2.6762
(2.4945)&
-4.6474***
(1.2306)&
-8.2323***
(1.0886)&
-1.4812
(4.4703)\\
$\mu_E$ &
1.8041
(2.0949)&
0.5473***
(0.1991)&
0.0335
(0.5346)&
2.0632
(3.3278)\\
$\eta$&
0.0145
(0.0135)&
0.0263**
(0.0125)&
-0.0054
(0.0245)&
0.0235
(0.0289)\\
		\hline
	\end{tabular}
	\caption{Maximum Likelihood Estimates for Knowledge Spillover (geographic distance)\label{table:MLE_geographic_distance}}
	\begin{tablenotes}
		\centering
		\small
         \item Signif. codes:  0 ‘***’ 0.001 ‘**’ 0.01 ‘*’ 0.05 ‘.’ 0.1 ‘ ’ 1
         \item (numbers in brackets are standard deviation of the regression coefficient)
	\end{tablenotes}
\end{table}

We see that, first of all, from a nationwide perspective, all factor inputs have a positive effect on regional new knowledge outputs (the number of authorized patents), and all variables pass significance test for $p$-value less than $0.1$. It fully shows that external knowledge spillovers actually contribute to regional innovative activities. Compared with R\&D capital reserve, R\&D personnel and exterior knowledge spillovers, local knowledge stock plays a more important role in regional new knowledge production, which indicates that regional innovative activities in China mainly rely on learning existing knowledge to create or produce new knowledge. Moreover, according to estimation results from Table \ref{table:MLE_geographic_distance}, the sum of elasticity of R\&D capital reserve, exterior knowledge spillovers and local knowledge stock to new knowledge output is less than one, that is to say $\beta_1+\beta_2+\beta_3<1$. It verifies the assumption we made for simplicity in the general innovative growth model, namely $\beta+\theta+\mu<1$, indicting that the situation of explosive knowledge growth does not exist in China. However, $\beta+\theta+\mu+\gamma>1$ reveals increasing return to scale of regional knowledge production, which corresponds with the development stage in China now. Coefficient $\eta>0$ shows that the level of inefficiency in regional knowledge productions in China decays over time.

Secondly, we analyze the estimation outputs in these three areas. Through comparative analysis of estimated results in eastern, central, western areas, we can see that the elasticity of R\&D capital reserve to knowledge outputs in both central and western areas are negative and not statistically significant. It shows that R\&D inputs in central and western areas slightly contribute to regional innovations. Since central and western areas lag behind eastern developed areas in terms of economic development and innovative capacity, it is difficult for central and western areas to attract private R\&D funds. Then most of the R\&D programs in central and eastern are funded by local government. To meet criteria set by central government on the surface, local government tends to blindly invest a large amount of money into inefficient R\&D programs without an effective and detailed plan. As a result, most of the R\&D expense in central and western areas lacks a reasonable goal and then is wasted. Moreover, key ideas in innovative activities is hard to acquire for central and east areas, and thus independent innovative capacities in these areas are relatively weak. Both factors lead to a lower marginal knowledge output of R\&D capital reserve in the process of knowledge productions. However, production elasticity of R\&D personnel to new knowledge output in both central and western areas are greater than that in eastern areas. On one hand, there is a much larger number of R\&D personnel in eastern areas than that in central and western areas. Because of decreasing return to scale, average knowledge output per person in eastern areas declines, which further causes smaller production elasticity in eastern areas. On the other hand, polices on attracting high quality talents from developed areas to central and western areas have taken effect, greatly improving the quality of R\&D personnel in central and western areas. Compared with eastern and western areas, central areas present more efficient utilization of current knowledge stock to produce new knowledge. 

Finally, the elasticity of knowledge spillovers to new knowledge outputs in both central and western areas is statistically significant while elasticity in eastern areas is not and is even negative in Table \ref{table:MLE_geographic_distance}. It shows that knowledge spillovers present ``competitiveness” rather than ``complementarity” in developed eastern coastal areas. Innovative activities in east areas are more inclined to be exclusive and different regions in eastern areas lack effective R\&D cooperative mechanism to produce complementary knowledge or share common knowledge, which constrains knowledge spillovers or diffusion among regions. Moreover, legal protection of intellectual property rights is stronger in eastern areas than that in central and western areas, which to some extent constrains large-scope knowledge spillovers among regions in eastern areas. Regions in central areas make full use of exterior knowledge spillovers in the process of new knowledge production. On one hand, regions in the central area are located in hinterland. It is relatively convenient for R\&D people in these areas to cooperate and communicate with each other among regions due to advantageous geographical locations and rapid developments of public transportation. Therefore, interactive R\&D cooperation and knowledge exchange become relatively frequent. On the other hand, based on the location advantage, central areas play an indispensable role in linking eastern and western areas in terms of economic radiation and knowledge spread. Through effectively absorbing and learning exterior knowledge stock from the eastern area, central area not only improves its innovative capacity but also further spreads new knowledge produced by eastern area to poor western area.

\section{Knowledge Spillover Range based on Spatial Distance}
\label{Sec:Knowledge_Spillover_Range}
\subsection{Division of Spatial Weighted Matrix and Remodeling}
According to the space-time diffusion theory, in spatial dimension, knowledge gradually spread from one original region or central region to other periphery regions. In most cases, the innovative center applies new produced knowledge into practice at first, and then peripheral regions begin to learn and absorb diffused new knowledge and take advantage of them. However, the utilization rate of new knowledge will decay as spatial diffusion distance rises. Since China is a vast country, new knowledge produced by one specific region is impossible to sprawl into every region, especially for the region far away from the birthplace of new knowledge. To measure the radiation range or scope of knowledge spillovers in China, we accurately divided previous spatial weighted matrix $W_g$ into five parts. Based on the spatial geographical distance, namely $0-1000$ kilometers, $1000-2000$ kilometers, $2000-3000$ kilometers, $3000-4000$ kilometers and above $4000$ kilometers, $W_g$ matrix is divided into five different weighted matrix, that is $W_1$, $W_2$, $W_3$, $W_4$, $W_5$. The entries in these matrixes are expressed as follows:
\begin{equation}
w_{1ij} =     \left\{ \begin{array}{rcl}
&w_{1ij} & \mbox{distance between region i and region j on and below 1000 kilometers }
\\ 
&0 & \mbox{distance between region i and region j beyond 1000 kilometers} 
\end{array}\right.
\end{equation}
where $w_{1ij}$ and $w_{ij}$ respectively represents element in matrix $W_1$ and $W_g$. Take Beijing for example, regions far away from Beijing below $1000$ kilometers includes Tianjin, Hebei, Shanxi, Inner Mongolia, Liaoning, Shandong, Henan. The elements of association degree between these regions and Beijing in matrix $W_1$ are counterparts in matrix $W_g$ and then are zero in matrix $W_2$, $W_3$, $W_4$ and $W_5$. Since the impact of exterior knowledge stock on regional innovative activities has been split in the light of spatial geographical distance, the stochastic frontier model is no long suitable for regression analysis. Here we use the traditional linear panel data model, that is:
$$Y_{i}(t)=\alpha_i+X_{i}(t)^T \beta+\epsilon_{i}(t),$$
where $Y_{i}(t)$ is a dependent variable (scalar), $X_{it}$ is a $k\times 1$ column vector (including $k$ regressors), $\beta$ is a $k\times 1$ column vector (including $k$ regression coefficients), $\alpha_i$ represents individual specific components, $\epsilon_{i}(t)$ represents the error term. Combining general innovative growth model with spatial distance division yields a panel data model as follow
\begin{equation}
\begin{split}
\ln \dot{A}_i(t)=&\beta_0+\beta_1 \ln K_{r,i}(t)+\beta_2 \ln L r_i(t)+\beta_3 \ln A_i(t)+\beta_4 W_1\ln A(t)+\beta_5 W_2\ln A(t)\\
&+\beta_6 W_3\ln A(t)+\beta_7 W_4\ln A(t)+\beta_8 W_5\ln A(t)+\alpha_i+\epsilon_{i}(t).
\end{split}
\end{equation}
To test the influence of spatial geographical distances, we generate five different models to be regressed, by removing knowledge spillover interactive term weighted by longest spatial distance between two regions in order until only $W_1\ln A(t)$ is left. That is, model $j\in \{1,2,3,4,5\}$ is given by
\begin{equation}
\begin{split}
\ln \dot{A}_i(t)=&\beta_0+\beta_1 \ln K_{r,i}(t)+\beta_2 \ln L r_i(t)+\beta_3 \ln A_i(t)+\sum_{k=1}^j \beta_{3+k} W_k\ln A(t)+\alpha_i+\epsilon_{i}(t).
\end{split}
\end{equation}

\subsection{Analysis of Spatial Knowledge Spillover Scope}
To decide whether the fixed effects model or random effects model is appropriate, we apply the Hausman chi-square test (\cite{hausman1978specification}) on panel data model . The specific forms of five models are determined by the Hausman test and Table \ref{table:Hausman_test} presents the results, where we can see that all five models should be fixed-effects panel data model. Compared with Table \ref{table:MLE_time_distance} and \ref{table:MLE_geographic_distance}, elasticity of R\&D capital stock, R\&D personnel, regional knowledge stock to knowledge output keep the same sign as before and their magnitude does not change greatly. Moreover, through observing estimates of five different knowledge spillover terms, it can be seen that basically only knowledge spillovers within $1000$ kilometers have a significant effect on regional innovation activities. Elasticity of exterior knowledge stock beyond $1000$ kilometers to regional knowledge output is not statistically significant. It fully shows that the radiation range of knowledge spillovers is within $1000$ kilometers. The knowledge spillover effect on regional innovative activities will decay fast when the distance between two regions is beyond $1000$ kilometers. For example, only little knowledge spreads from eastern areas to far western areas, thus making it difficult for regions in developing western areas to benefit from exterior knowledge stock in developed eastern areas. In general, most regions in central areas are located in radiation range of knowledge spillovers from eastern area, namely within $1000$ kilometers. For instance, knowledge stock in eastern areas (Tianjin, Hebei, Shanghai, Jiangsu, Zhejiang, Fujian, Shandong) have a great impact on Anhui’s innovative activities. It matches the conclusion drawn from estimated results in Table \ref{table:MLE_time_distance} and \ref{table:MLE_geographic_distance} that central areas have the highest utilization rate of exterior knowledge stock.
\begin{table}
	\centering
	\centering
	\setlength{\tabcolsep}{8pt}
	\small
	\begin{tabular}{cccccc}
		\hline\hline
		\thead{Hausman test} & \thead{Model 1} & \thead{Model 2} & \thead{Model 3} & \thead{Model 4} & \thead{Model 5}\\
		\hline
		Chi2(n) &
81.17&
107.94&
125.07&
91.65&
102.92\\
P-value&
0.00&
0.00&
0.00&
0.00&
0.00\\
		\hline
Conclusion& RNH &
RNH &
RNH &
RNH &
RNH \\
		\hline
	\end{tabular}
		\caption{Hausman test \label{table:Hausman_test}}
	\begin{tablenotes}
		\centering
		\small
		\item (rejecting null hypothesis (RNH) under the significant level of $1\%$)
	\end{tablenotes}	
\end{table}

\begin{table}
	\centering
	\centering
	\setlength{\tabcolsep}{8pt}
	\small
\resizebox{\textwidth}{!}{\begin{tabular}{cccccc}
		\hline\hline
		\thead{Explanatory\\ variables} & \thead{Model 1\\ Fixed effect} & \thead{Model 2\\ Fixed effect} & \thead{Model 3\\ Fixed effect} & \thead{Model 4\\ Fixed effect}& \thead{Model 5\\ Fixed effect}\\
		\hline	
		$\ln K r$&
0.2603***(0.0699)&
0.2442***(0.0735)&
0.1868**(0.0794)&
0.1889**(0.0795)&
0.1982**(0.0799)\\
$\ln L r$&
0.3497***(0.0581)&
0.3422***(0.0591)&
0.3334***(0.0591)&
0.3342***(0.0591)&
0.3216***(0.0601)\\
$\ln A$&
0.7509***(0.0529)&
0.7513***(0.0529)&
0.7605***(0.0529)&
0.7565***(0.0534)&
0.7138***(0.0642)\\
$w_1\ln A$&
0.3298***(0.0618)&
0.3033***(0.0625)&
0.3481***(0.0673)&
0.3527***(0.0642)&
0.3907***(0.0716)\\
$w_2\ln A$& &
0.1197(0.1676)&0.1076(0.1671)&0.1268(0.1706)&0.1691(0.1741)\\
$w_3\ln A$& & &0.5682*(0.3031)&0.3522(0.4836)&0.3513(0.4833)\\
$w_4\ln A$& & & &
0.2999(0.5228)&-0.0955(0.6177)\\
$w_5\ln A$&& & & &
2.6829(2.2364)\\
		\hline
$F$&
800.57&
639.53&
537.92&
460.11&
403.36\\
Prob$>F$&
0&
0&
0&
0&
0\\
$R^2$&
0.7882&
0.7871&
0.7786&
0.7819&
0.7805\\
		\hline
	\end{tabular}}
	\caption{Estimates of Knowledge Spillover Range}
	\begin{tablenotes}
		\centering
		\small
		\item (blank spaces in the table represent variables absent from this model)
	\end{tablenotes}
\end{table}

\section{Convergence Property of Regional Knowledge Growth Rate}
\label{Sec:Convergence_Property}
\subsection{Steady Growth Rate and Convergence Indicator}
Based on the regional knowledge diffusion theory and steady solutions derived from the general innovative growth model, it can be seen that the speed of producing new knowledge in different regions will converge in the long run. Simply put, the knowledge growth rate in different regions will tend toward the same steady point, that is:
$$g_A^{\ast}=\frac{\gamma+\beta}{1-\theta-\beta-\mu}n.$$
According to estimated results in table \ref{table:MLE_time_distance}, we can plug $\gamma=0.2418$, $\beta=0.0921$, $\theta=0.7477$, $\mu=0.1377$ into the above equation. Moreover, $n$ can be calculated by taking the geometric average of the total number of R\&D people between $2000$ and $2016$, that is $n=(Lr_{2016}/Lr_{2000})^{1/16}-1=0.843\%$. Then we get the value of the steady knowledge growth rate, namely $g_A^{\ast}=0.1252$.

To verify the conclusion drawn from derivations of the general innovative growth model, we need to calculate convergence indicators of the knowledge growth rate from $2001$ to $2016$. In the light of traditional statistical methods, by calculating standard deviations of observable variables among regions year by year, we analyze convergence conditions of observable variables. In this paper, the observable variable is the knowledge growth rate. If standard deviations declines year by year, it shows that the gap of knowledge growth rate among regions narrows down over time, i.e. convergence. If standard deviations rise year by year, knowledge growth rates among regions diverge or lack of the convergence tendency. Detailed formulas are 
\begin{equation}
\label{eq:growthrate}
g_{it}=\frac{\dot{A}_i(t-1)}{{A}_i(t-1)}=\frac{P_i(t)}{A_i(t-1)}, \quad \sigma_t=\sqrt{\frac{1}{n} \sum_{i=1}^n(g_{it}-\frac{1}{n}\sum_{i=1}^n g_{jt})^2 },
\end{equation}
%
where $t$ spans from $2001$ to $2016$, and $i$ represents $31$ different regions in China.

\subsection{Interpretation of Result}
\begin{figure}
	\centering
	\includegraphics[scale=0.7]{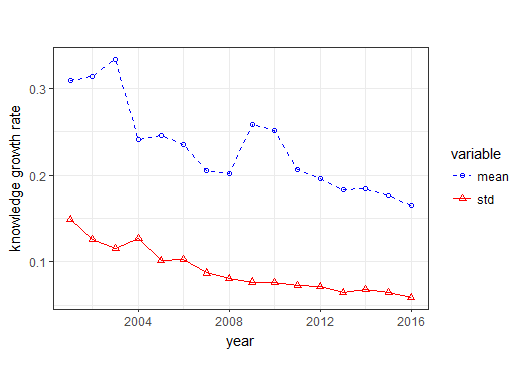}
	\caption{Variation Trend of Knowledge Growth Rate}
	\label{fig:plot}       
\end{figure}

By equation \eqref{eq:growthrate}, we calculate the mean and standard deviation of knowledge growth rates in all regions every year. From
Figure \ref{fig:plot}, we see that in general the standard deviations of knowledge growth rates among regions decline over time, which indicates that to some extent knowledge growth rates present the characteristic of convergence. It partially conforms to conclusions drawn from the theoretical model. Furthermore, the mean of knowledge growth rates among different regions shows the trend of shock going down over time and gradually tending toward the steady knowledge growth rate, that is $g_A^{\ast}=0.1252$. Current mean of knowledge growth rate is still higher than the steady one, showing that regions in China still have potentials in promoting knowledge growth rates within the next few decades.

\section{Conclusion and Suggested Policies}
\label{Sec:Conclusion}
\subsection{Conclusion}
In view of the theoretical and empirical analysis above, we conclude the paper as follows: 
\begin{enumerate}[(i)]
\item Innovative activities among regions in China to some extent present spatial association whose level have been gradually deepened over years. This confirms the existence of spatial knowledge spillovers among regions in China. 
\item From the analysis in the whole nation and in the middle of western areas, external knowledge spillovers actually promote regional innovative activities, and innovative activities in central areas benefit mostly from external knowledge stock. However, regions in developed eastern areas have not effectively turned external knowledge stock into the driving force of regional innovations. 
\item The radiation range or influencing scope of knowledge spillovers among regions in China is within $1000$ kilometers. The strength of the knowledge spillover effect on regional innovative activities will decline greatly when the distance between two regions is beyond $1000$ kilometers.
\item Based on derived solutions of the general innovative growth model, knowledge growth rates of different regions under knowledge spillovers will converge to the same steady-state growth rate in the long run. Empirical analysis shows that knowledge growth rates in China presents the characteristic of convergence over the past eleven years, which verifies derived solutions to some extent.
\end{enumerate}
	
In sum, innovative activities in China engender the spatial knowledge spillover effect. Though the main radiation range of knowledge spillover is within $1000$ kilometers, knowledge spillovers in fact promote regional innovative activities. Fortunately, developing regions benefit more from spatial knowledge spillovers than developed regions. As a result, the evolution principle of regional innovation without knowledge spillovers is the same as that with knowledge spillovers, namely convergence of regional knowledge growth rate.

\subsection{Policy Making Suggestion}
In the light of conclusions achieved, we propose the following suggested policies to achieve the goal of a sustainable and regionally balanced economic development:

\begin{enumerate}[(i)]
	\item The government should encourage knowledge sharing among different regions and optimize the external environment of knowledge spillovers. Owning to strong spatial association among regions in the process of knowledge production, government should make relevant policies, stimulating economic and innovative clustering central regions to transfer knowledge to periphery regions. The developing regions ought to search for opportunities to cooperate with and share knowledge with developed regions, which will not only improve their innovative capacities but also develop a mutually beneficial status. Moreover, poor regions should strengthen construction of network communication infrastructure and public transportation infrastructure such as airport, railway and harbor to compress space-time distance, so that they can absorb external knowledge spillovers more effectively.

	\item Regions with weak innovative capacity should promote regional learning abilities of knowledge spillovers to enhance the positive effect of knowledge spillovers on regional innovations. The government is further suggested to encourage private R\&D expense and knowledge accumulation, both of which contribute to increasing regional knowledge stock and then improving regional absorption and learning ability. Poor regions should make full use of public and private R\&D funds to attract R\&D talents with high qualities instead of simply introducing new knowledge. Moreover, knowledge spillovers in eastern areas present `competitiveness'. Therefore, the government should build up communication platforms for R\&D people who do similar kind of R\&D programs and companies ought to publish and apply for patents of R\&D achievements as soon as possible.

	\item R\&D institutions and companies should build up a cooperative network of innovative activities. In order to enlarge the radiation range of knowledge spillovers, based on this social cooperative network system, regional innovative activities in different regions should be associated accordingly. Local governments should communicate with each other and find a mutually beneficial equilibrium before making economic policies. Only by establishing a widespread social cooperative network, can developed regions achieve a sustainable and innovation-oriented economic development. The government should encourage R\&D people and inventors in developed regions to participate knowledge exchange targeted activies to communicate and present their R\&D achievements to local R\&D people.

	\item By allocating research funds to most potential R\&D programs, different regions can achieve a balanced development. Based on the widely agreed evolution rule of economic development, the ``invisible hand'’ in the competitive market will eventually yield optimal allocations and hinder increasing economic gaps among regions by allocating funds to projects with higher returns. Therefore, under a mature and market-oriented economic system, in order to promote regional independent innovations, government should stick to the law of the market and understand that the market is optimizing the limited resource allocation. The government should minimize relevant legal regulations about maintaining market order and avoid over intervening in economic activities. Only in this way, will social optimum be achieved.
\end{enumerate}

\bibliography{\jobname}

\end{document}